# Dual-polarization control of broadband nonreciprocal thermal radiation by combining local and nonlocal metasurfaces

Shuang Xia[1,2]†, Mengqi Liu[3,4]†, Wenjian Wan[5]†, Jialong Wang[6], Weihao Yang[1,2], Chaoran Wang[1,2], Huiqin Ma[6], Jun Qin[6], Hua Li[5]*, Yuan Wang[1,2], Lei Bi[6]*, Chengwei Qiu[4]* and Xiaobo Yin[1,2]*

*1 Department of Mechanical Engineering, The University of Hong Kong, Hong Kong 999077, China*

*2 State Key Laboratory of Optical Quantum Materials, The University of Hong Kong, Hong Kong 999077, China.*

*3 Institute of Engineering Thermophysics, MOE Key Laboratory for Power Machinery and Engineering, School of Mechanical Engineering, Shanghai Jiao Tong University, Shanghai 200240, China*

*4 Department of Electrical and Computer Engineering, National University of Singapore, Singapore 117583, Singapore*

*5 Key Laboratory of Terahertz Solid State Technology, Shanghai Institute of Microsystem and Information Technology, Chinese Academy of Sciences, Shanghai, 200050 China*

*6 National Engineering Research Center of Electromagnetic Radiation Control Materials, University of Electronic Science and Technology of China, Chengdu 610054, China*

** Email: hua.li@mail.sim.ac.cn; bilei@uestc.edu.cn; eleqc@nus.edu.sg; xbyin@hku.hk*

*† These authors contributed equally to this work.*

## ABSTRACT:

Nonreciprocal thermal radiation offers a route to decouple spectral directional absorptivity and emissivity, thereby enabling new paradigms in thermal-photonic systems. However, in magneto-optical platforms, the intrinsic gyroelectric response generally confines observable nonreciprocity to transverse-magnetic (TM) polarization, while the transverse-electric (TE) response is absent. In this work, we experimentally demonstrate, for the first time, a local thermal metasurface strategy to activate TE-polarized nonreciprocity by creating artificial gyromagnetic response in a gyroelectric semiconductor platform. We further extend this mechanism to broadband dual-polarization operation employing a nonlocal thermal metasurface, which combines a resonator supercell with gradient-doped epsilon-near-zero magneto-optical multilayers. Pronounced absorptivity contrast is maintained over 22–27 μm for TE polarization and 19–27 μm for TM polarization. This platform provides a mechanism-based route to achieve broadband and dual-polarization nonreciprocal thermal absorption, opening new opportunities for advancing radiative energy-conversion devices.

## INTRODUCTION

Nonreciprocity describes a set of fundamentally interesting phenomena in photonics[1–3]. By employing, for example, magneto-optical materials, Lorentz reciprocity can be broken and thereby enabling a wide range of important applications. More recently, nonreciprocity has been extended to the realm of thermal radiation[4,5]. It has attracted significant interest because of its potential to violate the fundamental Kirchhoff's law of thermal radiation[6], which states that the absorptivity and emissivity of a material are identical at the same temperature, wavelength, angle, and polarization. In nonreciprocal systems, this equality no longer holds[7,8]. This capability opens new opportunities for next-generation photonic energy systems and offers a potential route toward exceeding the previously established limit of conventional reciprocal thermal energy harvesting technologies[9], including photovoltaics, thermophotovoltaics[10] and solar energy conversion[11].

Over the past decades, extensive theoretical work has proposed potential designs to realize near-complete violation of Kirchhoff's law in nonreciprocal systems[12–14]. Recently, several experimental demonstrations based on magneto-optical materials have been reported. Shayegan et al. fabricated a guided-mode resonance (GMR) waveguide structure patterned on an InAs slab[15]. The spectral directional emissivity and absorptivity were measured separately, which offers direct experimental evidence for the violation of Kirchhoff's law in thermal radiation. But the operating bandwidth in most designs remained limited owing to the inherently narrowband characteristic of the resonance mode. To overcome this challenge, Liu et al. ingeniously harnessed the epsilon-near-zero (ENZ) properties of InAs material[16]. Nonreciprocal absorption spectra over a broad mid-infrared band were demonstrated using InAs multilayers with gradient doping concentration. In parallel, nonreciprocal properties as a function of scattering angle and under high temperature were systematically investigated[17,18]. More recently, strong nonreciprocal thermal emission was experimentally observed, with the difference between emissivity and absorptivity reaching as high as 0.43 under 5 T magnetic field[19].

However, as illustrated in Fig. 1a,b, almost all experimental demonstrations based on magneto-optical film can only exhibit nonreciprocal thermal radiation under TM polarization, whereas the TE-polarized nonreciprocal response is essentially absent. This polarization constraint originates not from the geometry design, but comes from the inherent gyroelectric property of natural magneto-optical materials[20,21]. Lifting the polarization constraint plays a vital role in nonreciprocal thermal photonics, since the thermal radiation is typically unpolarized and such a fundamental limitation will effectively halve the theoretical maximum efficiency of energy conversion systems. Although several theoretical works have suggested possible routes toward TE-polarized nonreciprocal thermal emission[22–24], experimental demonstration in intrinsically gyroelectric systems, especially while sustaining broadband and dual-polarized nonreciprocity, remains unresolved. The primary challenge is to induce gyromagnetic response in intrinsically gyroelectric magneto-optical materials, thereby allowing previously inaccessible TE-polarized waves to exhibit nonreciprocity. Beyond this polarization activation, achieving broadband operation imposes a more stringent requirement as the localized magnetic resonances that enable gyromagnetic response are typically narrowband, making it challenging to simultaneously achieve broadband response, polarization insensitivity and strong nonreciprocal contrast in a single structure.

Here we firstly demonstrate the breaking of polarization constraint in nonreciprocal thermal radiation enabled by unique magnetic-dipole (MD) resonance in local metasurface, as shown in Fig. 1c**.** This Mie-type mode acts as an out-of-plane magnetic dipole, which can reconfigure the local electric field orientation and engineer the gyromagnetic permeability tensor. Consequently, even under TE polarization, the magneto-optical medium can still interact with the incident light, thereby enabling observable nonreciprocal response. Furthermore, to overcome the narrowband limitation of a single localized resonance, we extend the design to a collectively coupled nonlocal metasurface. As shown in Fig. 1d, size-varied resonators are arranged into a supercell and interact with gradient doped ENZ magneto-optical multilayers to enable broadband nonreciprocal property. The detailed geometric parameters and doping concentrations were determined through genetic algorithm method to achieve an optimal balance, thereby simultaneously ensuring polarization insensitivity, broadband response and strong nonreciprocity. Our demonstration introduces a new degree of freedom for manipulating nonreciprocal thermal radiation and opens a new avenue for the development of the next-generation photonic energy systems.

# RESULTS AND DISCUSSION

## Mechanism of breaking the polarization constraint

As discussed below, a bare magneto-optical (MO) film exhibits nonreciprocal response only under TM polarization. For the n-type doped InAs film, when an external magnetic field is applied along the *y*-direction which is in the sample plane but perpendicular to the plane of incidence, the permittivity tensor becomes asymmetric with nonzero off-diagonal elements $\varepsilon = \left[\varepsilon_{xx}, 0, \varepsilon_{xz}; 0, \varepsilon_{yy}, 0; \varepsilon_{zx}, 0, \varepsilon_{zz}\right]$. According to the Drude model, this off-diagonal term can be expressed as: $\varepsilon_{xz} = -\varepsilon_{zx} = i \cdot \omega_p^2 \omega_c / \omega\left[\left(\omega + i\Gamma\right)^2 - \omega_c^2\right]$, where $\omega_p = \sqrt{ne^2 / (m_e \varepsilon_0)}$ and $\omega_c = eB / m_e$ are the plasma and cyclotron frequencies, $\Gamma = e / \left(m^* \mu\right)$ is relaxation rate. The dielectric displacement vector **D** and the electric field vector **E** In this configuration can be derived as[25,26]:

$$\mathbf{D} = \varepsilon_d \mathbf{E} + i[g_y \times \mathbf{E}] \tag{1}$$

Here $\varepsilon_d$ is the diagonal permittivity and $g_y$ is the gyration vector, which is proportional to the magnetization $M_y$. Under TE polarization, the incident electric field has only the *y*-component which is parallel to the external magnetic field. As a result, it cannot effectively couple to the off-diagonal terms ( $[g_y \times \mathbf{E}] = 0$ ). This behavior fundamentally originates from the intrinsic gyroelectric property of conventional magneto-optical materials. Fig. 2a shows the calculated absorption spectra of a bare InAs film under magnetic fields of ±1.5 T. The film has a doping concentration of $7\times10^{23}\,\mathrm{m}^{-3}$ and a mobility of $9000\,\mathrm{cm}^2\,\mathrm{V}^{-1}\,\mathrm{s}^{-1}$, which are typical for experimentally grown films. And the strength of nonreciprocal response in thermal radiation can be quantified by the contrast between absorptivity and emissivity: $\eta = \alpha\left(\theta, \lambda, B\right) - e\left(\theta, \lambda, B\right)$. Since no diffraction channel is supported in our structures, the directional emissivity should be equal to the absorptivity of the opposite angular channel[13,27–29]: $e\left(\theta, \lambda, B\right) = \alpha(-\theta, \lambda, B)$. Besides, the mirror-symmetry with respect to the y-z plane leads to the equation [16,19] $\alpha(-\theta, \lambda, B) = \alpha(\theta, \lambda, -B)$. Thus, the nonreciprocity could be equivalently evaluated from the absorptivity contrast under magnetic-field reversal: $\eta = \alpha\left(\theta, \lambda, B\right) - \alpha\left(\theta, \lambda, -B\right)$. Consistent with the theoretical prediction, Berreman resonance is excited under TM polarization, resulting in a distinct absorption peak at around λ=24 μm. The absorption spectra under opposite magnetic field show clear contrast, indicating strong nonreciprocal behavior. In contrast, under TE polarization, the Berreman mode cannot be supported and the absorptivity remains unchanged when magnetic field reversal, as shown by the red curves. These results indicate that the nonreciprocal functionality is essentially unavailable under TE polarization.

To overcome the intrinsic polarization restriction in nonreciprocal thermal radiation, we propose the local metasurface patterned on top of the magnetized InAs layer. The metasurface is composed of periodic germanium microdisks. Owing to its high refractive index and low optical loss in the infrared range, Ge is particularly suitable for constructing dielectric resonators that support strong Mie-type resonances. Multipole decomposition in Fig. 2b shows that the dominant contribution arises from a MD resonance (see calculation details in section S2). This unique mode is characterized by a circulating electric-field vector in the $x$-$y$ plane, corresponding to an effective magnetic dipole moment oriented along the out-of-plane $z$-direction[30,31]. Such a near-field behavior is crucial for TE polarized ($E_y$, $H_x$, $H_z$) nonreciprocal response. In the bare MO film, TE-polarized incident light contains only the $E_y$ component and cannot couple to the off-diagonal permittivity components ($\varepsilon_{xz}$ and $\varepsilon_{zx}$). By contrast, the MD resonance reconfigures the local field distribution and introduces an additional $E_x$ component that is absent in the bare film, enabling effective interaction with MO medium. In comparison, electric dipole (ED) at longer wavelength mode exhibits a fundamentally different field distribution, with the electric field oscillating predominantly along the $y$- direction and behaving as an in-plane electric dipole. Thus, this mode will result in negligible nonreciprocal response under TE polarization. Fig. 2e and 2f present the calculated angle-resolved nonreciprocal absorption spectra of this hybrid metasurface under different polarizations. For TM-polarized light ($H_y$, $E_x$, $E_z$)**,** both the ED and MD resonances can enhance the nonreciprocal contrast when strong field confinement is achieved in the MO layer. In contrast, for TE polarization, only the MD resonance can enhance the nonreciprocal response, whereas the contribution from the ED mode is nearly negligible. Another notable feature is the weak angular dispersion of these Mie-type resonances, allowing pronounced nonreciprocal contrast to be sustained even at small incident angles. This behavior differs fundamentally from other works based on the Berreman mode[32,33], which requires a large oblique incidence. Interestingly, these resonance modes exhibit opposite signs of $\eta$ under TM polarization. Fig. 2g further extracts the nonreciprocal absorption contrast $\eta$ at an incident angle of 45°. Under TM polarization, $\eta$ exhibits a clear sign reversal between the ED and MD resonances, indicating that these two modes contribute differently to the nonreciprocal response. According to temporal coupled-mode theory[13,34], the sign of the nonreciprocal contrast is determined by the relative magnitudes of the in-coupling and out-coupling rates of the resonant mode. Previously, the reversal of nonreciprocal contrast was achieved by reversing the external magnetic field direction. In contrast, our results show that the sign of $\eta$ can be regulated by the type of excited resonance mode under a fixed magnetic bias. This control of the nonreciprocal sign through resonant-mode selection provides an additional degree of freedom for engineering nonreciprocal thermal radiation.

## Experimental observation of TE-polarized nonreciprocal absorption

We fabricated the hybrid MO metasurface consisting of a periodic array of Ge microdisks patterned on 700-nm-thick InAs film grown on n-doped InAs substrate **(**see Methods for details**)**. A scanning electron microscopy image of the sample is shown in Fig. 3a. The angle-resolved absorption spectrum without an external magnetic field were first characterized using the Fourier transform infrared spectrometer (FTIR) equipped with a variable-angle reflection accessory and linear polarizers. As shown in Fig. 3c and 3f**,** the measured spectra exhibit similar characteristics under both TE and TM incidence. Two distinct absorption peaks[35] wavelengths show weak dependence on the incident angle. Numerical simulations calculated using CST are shown in Supplementary Information (fig. S3), which are in good agreement with the experimental results. The nonreciprocal absorption spectra at $\theta$=45° were then measured by a custom-built infrared MO Kerr effect characterization setup (shown in Fig. 3b). This angle was chosen as a representative oblique incidence condition used in TMOKE measurements, where the measurable MO response is expected[35]. The detailed angle-resolved absorption spectra showing clear angular asymmetry are presented in the Supplementary Information (fig. S5). Under TM polarization, a strong contrast of the absorptivity is observed near both the ED and MD resonances, with the maximum $\eta$ reaching 0.23 at $\lambda$=27 μm. More importantly, under TE polarization, a clear absorptivity contrast is also resolved, but only in the vicinity of the MD resonance. In sharp contrast, the measurement results of bare InAs film show a strong nonreciprocal behavior only under TM polarization but negligible modulation under TE polarization, as presented in the Supplementary Note 4. The experimental spectra agree well with the simulation results (fig. S4). The measured absorption peaks are broader and exhibit slight shifts of the resonance wavelength, which can be attributed to fabrication tolerance, surface roughness, intrinsic material loss, and experimental noise. We further measured the magnetic field dependence of the nonreciprocal response, as shown in Fig. 3e and 3h. The absorption contrast increases approximately linearly with the magnetic field under both polarizations. This behavior originates from the MO property of InAs, whose off-diagonal permittivity tensor element can be expressed as $\varepsilon_{xz} = -\varepsilon_{zx} = i\omega_p^2\omega_c / (\omega\left[(\omega + i\Gamma)^2 - \omega_c^2\right])$, where $\omega_c = eB / m_e$ is the cyclotron frequency. Therefore, the off-diagonal element is proportional to the magnetic field. These results clearly confirm the nonreciprocal response of the proposed metasurface[36,37]. To further clarify the origin of the observed TE-polarized nonreciprocal absorption, we calculated the effective gyrotropic tensors of the metasurface. Basically, the permittivity and permeability tensors of gyroelectric and gyromagnetic medium take the form:

$$\overline{\overline{\varepsilon}} = \begin{bmatrix} \varepsilon_x & 0 & -i\cdot\gamma \\ 0 & \varepsilon & 0 \\ i\cdot\gamma & 0 & \varepsilon \end{bmatrix}, \overline{\overline{\mu}} = \begin{bmatrix} \mu_x & 0 & -i\cdot\kappa \\ 0 & \mu & 0 \\ i\cdot\kappa & 0 & \mu \end{bmatrix}, \tag{2}$$

where $\varepsilon$ and $\mu$ are the diagonal components, $\gamma$ and $\kappa$ are the off-diagonal components of permittivity and permeability. The corresponding eigenmode relations under different polarization could be further expressed as[20]:

$$\begin{vmatrix} \varepsilon_x & \frac{k}{k_0}cos\phi & \frac{k}{k_0}sin\phi \\ \frac{k}{k_0}cos\phi & \mu & -i\cdot\kappa \\ \frac{k}{k_0}sin\phi & i\cdot\kappa & \mu \end{vmatrix} = 0\ (TE), \begin{vmatrix} \mu_x & -\frac{k}{k_0}cos\phi & -\frac{k}{k_0}sin\phi \\ -\frac{k}{k_0}cos\phi & \varepsilon & -i\cdot\gamma \\ -\frac{k}{k_0}sin\phi & i\cdot\gamma & \varepsilon \end{vmatrix} = 0\ (TM) \tag{3}$$

These equations indicate that the TE-polarized response is governed by the off-diagonal permeability coefficient κ, whereas that for TM polarization is governed by the off-diagonal permittivity coefficient γ. Therefore, in conventional gyroelectric MO films, where γ is non-zero but κ is negligible, nonreciprocal response is observed only under TM polarization. However, when gyromagnetic response is introduced, the magnetic field components can couple to the off-diagonal permeability tensor, thereby enabling the TE-polarized nonreciprocal response[38,39]. Equivalently, the generalized Cotton–Mouton relations can be expressed as[40]:

$$\begin{aligned} n^2_{TE} &= \varepsilon_x \mu \left(1 - Q^2_{\ m}\right),\ Q_m = \kappa / \mu \\ n^2_{TM} &= \varepsilon \mu_x \left(1 - Q^2\right),\ Q = \gamma / \varepsilon \end{aligned} \tag{4}$$

The hybrid structure composed of local metasurface and MO film were treated as a layer of equivalent material, as shown in Fig. 3**i**. Then we used the 4 × 4 transfer matrix method (TMM) to retrieve the effective permittivity and permeability tensors. The off-diagonal elements were obtained by fitting the simulated amplitude and phase difference of the reflection coefficient between positive and negative magnetic fields, with details provided in Supplementary Note 5. Fig. 3j,k show the real and imaginary parts of $Q_m$ and $Q$ for TE and TM polarization, respectively. The results confirm that our proposed metasurface exhibits an artificial gyromagnetic response under TE polarization, while retaining the conventional gyroelectric response under TM polarization.

## Broadband & dual-polarization nonreciprocal thermal absorption

Next, we further demonstrate that the proposed metasurface can be extended to broadband nonreciprocal absorption while preserving dual-polarization functionality. Although the results in Figs. 2 and 3 show that the MD resonance can lift the intrinsic polarization restriction, the nonreciprocal response is still limited to a narrow spectral range due to the strongly localized resonant nature of the MD mode. Previous studies have shown that gradient-doped InAs multilayers can realize broadband nonreciprocal absorption[16], but such platforms cannot work under TE polarization. Therefore, achieving broadband and dual-polarization nonreciprocity in a single device remains a nontrivial challenge. To overcome this limitation, we design a nonlocal metasurface composed of multi-dielectric-resonator supercell integrated with gradient-doped InAs magnetized films, as shown in Fig. 4a. The InAs multilayer will ensure strong MO effect over a wide band, whereas collective interactions among the different resonators within each supercell give rise to spectrally distributed MD resonant modes[41,42]. Their hybridization enables broadband nonreciprocal absorption under both TE and TM polarizations.

The dielectric metasurface is designed as a 2×2 supercell with a period P=4×p, where p is the period of the structure in Fig. 3. The height of the resonators is fixed at $t_1$=2 μm. In the single resonator structure, a circular disk is used as a simple model to demonstrate TE-polarized nonreciprocity through a localized MD resonance. For the broadband design, the circular resonator is generalized to an elliptical geometry to provide additional degrees of freedom. Each supercell contains four Ge elliptical microdisks with independently tunable dimensions. The long and short axes of each microdisk are treated as separate optimization parameters, denoted by [$R_{L1}$, $R_{L2}$, $R_{L3}$, $R_{L4}$] and [$R_{s1}$ $R_{s2}$, $R_{s3}$, $R_{s4}$]. The resonance wavelength of MD mode is strongly correlated with the particle size and can be approximately estimated from the equation : $2R \approx \lambda / n_{particle}$ [30]. This size-dependent resonance behavior allows us to directly tune the target wavelength by optimizing these geometrical parameters. Notably, fourfold rotational symmetry (C4 symmetry) is imposed during the optimization to minimize polarization sensitivity introduced by the geometrical asymmetry. The bottom part consists of a 5-layer gradient doped InAs film, where each layer has a fixed thickness of $t_2$=300 nm. The carrier concentrations of the individual layers are also taken as optimization variables to control the ENZ wavelengths and obtain magneto-optical response. However, since both material and geometrical parameters are tunable and the coupling among different resonant elements is complicated, the design of a high-performance structure becomes highly challenging. We employ a genetic algorithm method (GA) to optimize these parameters and maximize the nonreciprocal absorptivity contrast over the 22–26 μm range while preserving polarization-insensitive operation. Each population member is encoded as a binary chromosome. The optimization target is evaluated by the following fitness function:

$$\text{cost} = \sum_{\lambda_i} {}_{TE/TM} \left(1-\eta_{\theta,B}\right) \tag{5}$$

where $\eta_{\theta,B}$ represents the intensity of nonreciprocity and $\lambda_i$ are the target wavelengths. Details of the GA method and the optimization histories can be seen in the Supplementary Note 6.

Fig. 4d presents a schematic of the GA-optimized broadband device and a SEM image of the fabricated sample. The white dashed outline marks one supercell, which consists of four resonant elements with different sizes arranged in a C4-symmetric configuration. The structure is integrated on multilayer InAs films whose carrier concentration decreases monotonically from the bottom to the top layer. Detailed optimized geometric and material parameters are summarized in Tables 1 and 2.

| | **ne1** | **ne2** | **ne3** | **ne4** | **ne5** |
|---|---|---|---|---|---|
| doping concentration ($\times 10^{17}$ cm$^{-3}$) | 5.24 | 5.63 | 5.75 | 5.88 | 8.05 |

**Table 1. Optimized doping concentration of each InAs films**

| | **Radius1** | **Radius2** | **Radius3** | **Radius4** |
|---|---|---|---|---|
| Rx (μm) | 3.19 | 1.82 | 2.14 | 3.03 |
| Rx (μm) | 5.07 | 1.74 | 1.85 | 3.63 |

**Table 2. Optimized geometric parameter of each Ge resonator**

As shown in Fig. 4 e and f, the measured angle-resolved absorption spectra at B=0 exhibits strong absorption over a broad spectral and angular ranges, in clear contrast to the result of single resonator patterned on a single InAs layer. More importantly, the spectra under TE and TM polarizations remain highly consistent. Although slight shifts of resonance wavelength can be identified, the simultaneously broadband and polarization-independent absorption characteristics are well preserved. The near-field distributions in Supplementary Information (fig. S15) further clarify the role of strongly interacting resonators within the supercell. At different wavelengths, the field enhancement is redistributed across distinct resonators, indicating that the broadband response arises from spatially extended resonance modes rather than independent local resonances (fig. S14). The collective coupling among size-varied resonators contributes to the broadband absorption under both TE and TM polarizations within the optimized frequency band. We next measured the nonreciprocal absorption spectra at an incidence angle of 45° (Fig. 4 g–h). The optimized device exhibits pronounced absorptivity contrast under both TM and TE polarizations throughout the target wavelength regime. For TE polarization, the nonreciprocal response is maintained over **22–27 μm** and the maximum value of $\eta$ reaches **0.03,** which is larger than that of the single resonance device shown in Fig. 3. For TM polarization, the operating band is further extended to **19–27 μm**. This broader bandwidth is attributed to the additional contribution from the MO response of the n-doped InAs substrate, whose ENZ wavelength is in the vicinity of **18–19 μm**. All measured spectra are in good agreement with the numerical results (fig. S11). To further assess the angular dependence of the proposed nonreciprocal device, we also calculated the polar plots of the absorptivity contrast averaged over the target wavelength range, as presented in the Supplementary Information (fig. S13). It should be emphasized that the sign of the nonreciprocal contrast $\eta$ also remains consistent throughout the optimized frequency band under both TE and TM polarizations. This feature is critical for broadband, polarization-insensitive thermal photonic applications. As discussed above, different resonances will exhibit opposite signs of nonreciprocal contrast because of their distinct coupling with MO material. This means that the preferred emission directions could be reversed, leading to the cancellation of the net radiation and reducing the efficiency of energy-harvesting devices. In our optimized structure, the consistent sign of $\eta$ can ensure that the broadband response contributes constructively to the net emissivity–absorptivity contrast.

Optimizing the device for dual-polarization operation will introduce a trade-off between the TE and TM responses. Therefore, the TM-polarized nonreciprocal performance of this sample is not fully maximized. To further verify the flexibility of our design strategy, we also applied the same GA framework to optimize the broadband response only under TM polarization. The fabricated sample and the corresponding experimental results are shown in fig. S18. Compared with the dual-polarization optimized device, this sample exhibits a smoother and stronger nonreciprocal absorption spectrum over the **18–26 μm** range under TM polarization. Compared with previous work employing bare multi-layer InAs films, where 10 μm bandwidth was achieved using 14 layers, this device reaches a comparable bandwidth of nearly 8 μm using only 5 InAs layers by incorporating an artificially engineered nonlocal metasurface into the multilayer structure. This design offers a promising route for more compact and lightweight nonreciprocal thermal photonic devices. These results indicates that the proposed optimization framework can be readily adapted to different target functionalities, including polarization robust broadband operation and enhanced single polarization performance.

# CONCLUSION

In summary, we experimentally demonstrate that the polarization limitation in nonreciprocal thermal radiation can be overcome by our proposed MO metasurface. The strong MD mode excited in high-index dielectric resonator reconfigures the local electric field orientation and enable effective interaction between MO medium and TE-polarized light, which is absent in conventional platforms. This mechanism breaks the intrinsic polarization restriction and introduces an additional degree of freedom for engineering nonreciprocal thermal radiation, pushing nonreciprocal energy-conversion systems closer to their theoretical maximum efficiency. Furthermore, using a genetic algorithm, we proposed the hybrid nonlocal metasurface combing with multilayer ENZ MO films to realize broadband nonreciprocal absorption under dual polarization. The nonreciprocal absorption spectrum can be observed under both TE and TM polarizations across **22–27** μm, spanning nearly **5 μm** in bandwidth. For TM polarization, the nonreciprocal response is stronger and the operating bandwidth can be further extended to **19–27** μm. The geometric parameters of the dielectric metasurface and doping concentration of the ENZ layers provide a versatile platform to simultaneously manipulate the spectral bandwidth and the polarization dependence of the nonreciprocal thermal radiation. The revealed mechanism and the optimization framework proposed here can be also applicable to a broad range of MO material systems, including III–V semiconductors[43] and magnetized Weyl semimetals[44,45]. This hybrid MO metasurface platform opens a realistic pathway toward integrated nonreciprocal thermal-photonic devices for advanced energy harvesting and radiative energy management.

# MATERIALS AND METHODS

## Device fabrication

Magneto-optical (MO) InAs films with different doping concentrations were grown on n-type InAs wafers by molecular beam epitaxy (MBE). The substrates were S-doped (100) InAs with a carrier concentration of ~$2.53 \times 10^{18}$ $cm^{-3}$ and a thickness of 500 μm. To account for small deviations between the designed and the actual carrier concentrations caused by growth imperfections, an experimental calibration procedure was performed. Specifically, the measured infrared reflection spectrum of a single-layer InAs film was fitted with simulated results which were calculated using different doping concentrations, from which a scaling factor was extracted. This factor was then applied to estimate the actual carrier concentration of each InAs layer in the gradient-doped multilayer samples. The spectra calculated using these corrected doping concentrations showed good agreement with the experimentally measured results **(fig. S2)**.

After the growth of the InAs films, a 2-μm-thick germanium layer was deposited by electron-beam evaporation. The refractive index of Ge was obtained by fitting spectroscopic ellipsometry data (shown in **fig. S1**). Then the photoresist layer (AZ4620) with a thickness of 5 μm was spin-coated onto the sample surface. The resonator array was fabricated by laser direct writing according to the designed geometric parameters and subsequently transferred into the underlying Ge layer by inductively coupled plasma (ICP) etching. $SF_6$ and $O_2$ gases were used as the etching gases with a flow ratio of 1:4. The RF and ICP powers were set to 50 W and 100 W, respectively. After etching, the residual photoresist was removed by immersing the sample in N-methyl-2-pyrrolidone (NMP).

## Numerical Simulations and the GA optimization method

The multipole decomposition of Mie resonances in the hybrid MO metasurface was calculated using COMSOL MUITIPHYSICS (**see Supplementary Note 2 for details**). The absorption spectra and nonreciprocal responses of the structure were simulated using CST Studio Suite based on the finite integration technique, which provides higher efficiency than COMSOL for S-parameter calculations. Periodic boundary conditions were applied in both x and y directions for the unit cell. The permittivity of InAs was defined using Drude mode and the optical constants of Ge were taken from ellipsometry measurements. In the section of genetic algorithm optimization, MATLAB was interfaced with CST Studio Suite. The S-parameters obtained from CST were used to calculate the objective function in MATLAB, and the updated structural parameters generated by the algorithm were then imported into CST for the next iteration. **The optimization workflow is shown in Supplementary Note 7.**

## Sample Measurement

The angle-resolved absorption spectra without magnetic field were measured using Fourier-transform infrared spectroscopy (Bruker TENSOR II) equipped with an angle specular reflection accessory (Pike VeeMAX III). The angle of incidence could be varied from 30° to 80° in 1° increments. The accessory also includes a built-in polarizer mount, allowing selection of the polarization state of the incident light.

The nonreciprocal absorption spectra were measured using a custom-built infrared MO effect characterization setup **(Fig. 3 B).** The broadband infrared source in the FTIR spectrometer was guided through the internal mirrors and interferometer and then directed out of the instrument by a planar mirror. The beam was subsequently redirected by a silver-coated 90° off-axis parabolic mirror. A pair of 45° off-axis parabolic mirrors with a focal length of 10 inches was used to focus the beam onto the sample surface and to collect the reflected light, respectively. The reflected beam was then focused to the DTGS detector by another 90° off-axis parabolic mirror. A static magnetic field was applied parallel to the sample surface and perpendicular to the incident plane using a custom-built electromagnet with a maximum applied magnetic field up to 1.5 T. The magnetic field strength was controlled by the applied current and monitored using a gaussmeter placed close to the sample. For background calibration, the reflectivity of a silicon wafer coated with a 200-nm-thick gold film was first measured at the same position and the reflectivity spectra of the samples were normalized to this reference. Each reflection spectrum shown in this work was obtained by averaging three consecutive measurements on the same sample. To improve the signal-to-noise ratio, an integration time of 150 s was used for each measurement.

## ACKNOWLEDGMENTS

**Author contributions:** S.X. and C.W. performed the numerical simulations and the GA optimization. W.Y. performed the calculations of effective permittivity and permeability tensors. S.X., H.M., W.W., and H.L. contributed to film growth and device fabrication. J.W. and S.X. built the experimental characterization setup and performed the optical measurements. S.X. and M.L. wrote and revised the manuscript. L.B., C.Q., and X.Y. supervised the project and contributed to data analysis and interpretation. All authors discussed the results and commented on the manuscript.

**Competing interests:** The authors declare that they have no competing interests.

**Data and materials availability:** All data needed to evaluate the conclusions in the paper are present in the paper and/or the Supplementary Materials.

## Figures:

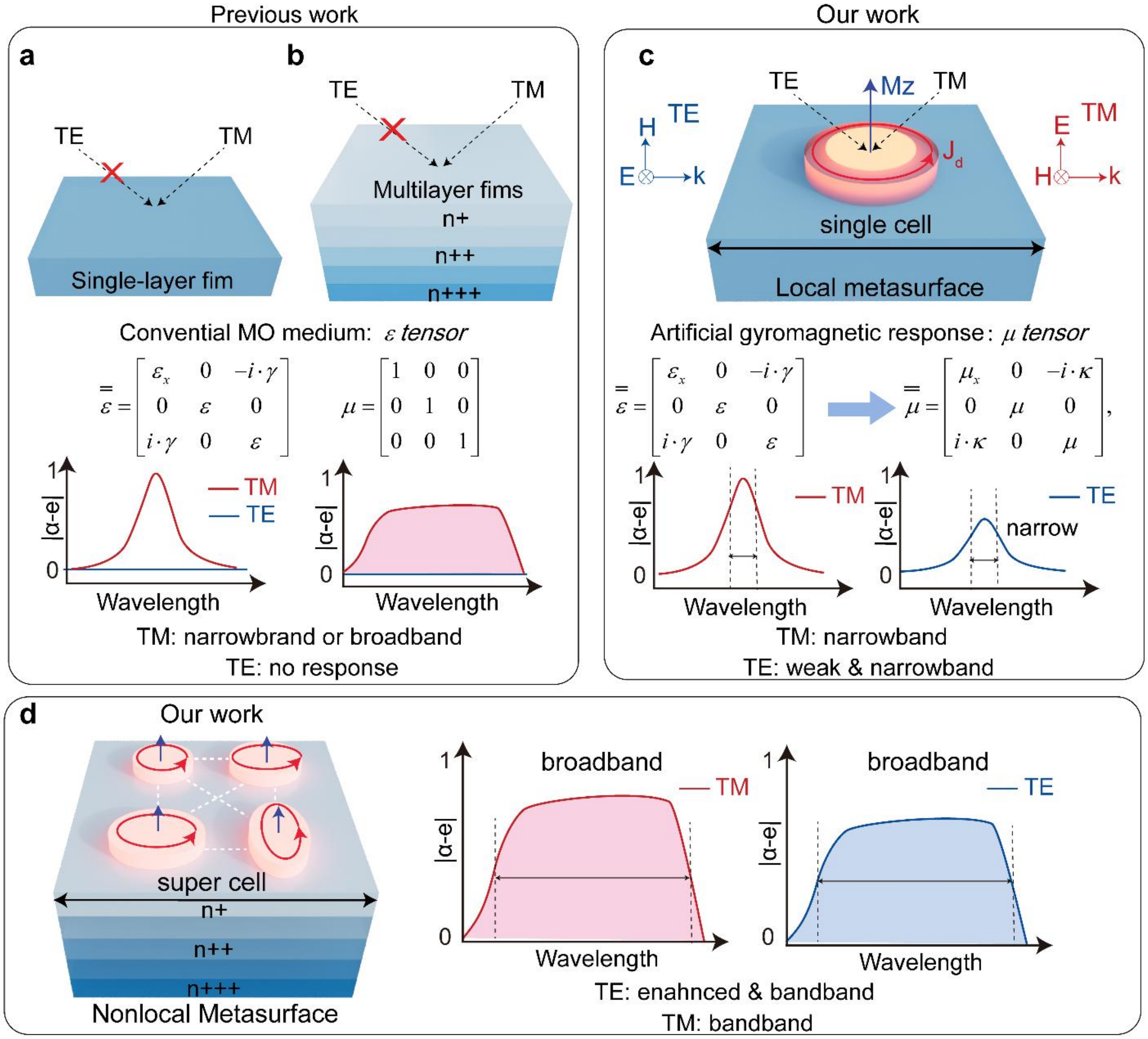

**Fig. 1. Principle of the hybrid metasurface realizing broadband nonreciprocal thermal radiation under both TE and TM polarizations. a, b,** Previous works based on single layer or multilayer films. Their nonreciprocal response originates from the intrinsic gyroelectric nature of conventional MO materials, whereas the permeability remains scalar (μ=1). This confines nonreciprocal thermal radiation to TM polarization, leaving the TE-polarized response essentially absent. **c,** Our work overcoming this TE polarization restriction by introducing magnetic-dipole mode in local metasurfaces. This unique resonance is characterized by a circulating displacement current in a single resonator and will create artificial gyromagnetic response within a narrow spectral range. **d,** Our proposed nonlocal metasurface combined with gradient-doped multilayers. Collective coupling among different resonators in the supercell will further enhance the TE response and extends dual-polarization nonreciprocity to a broadband regime.

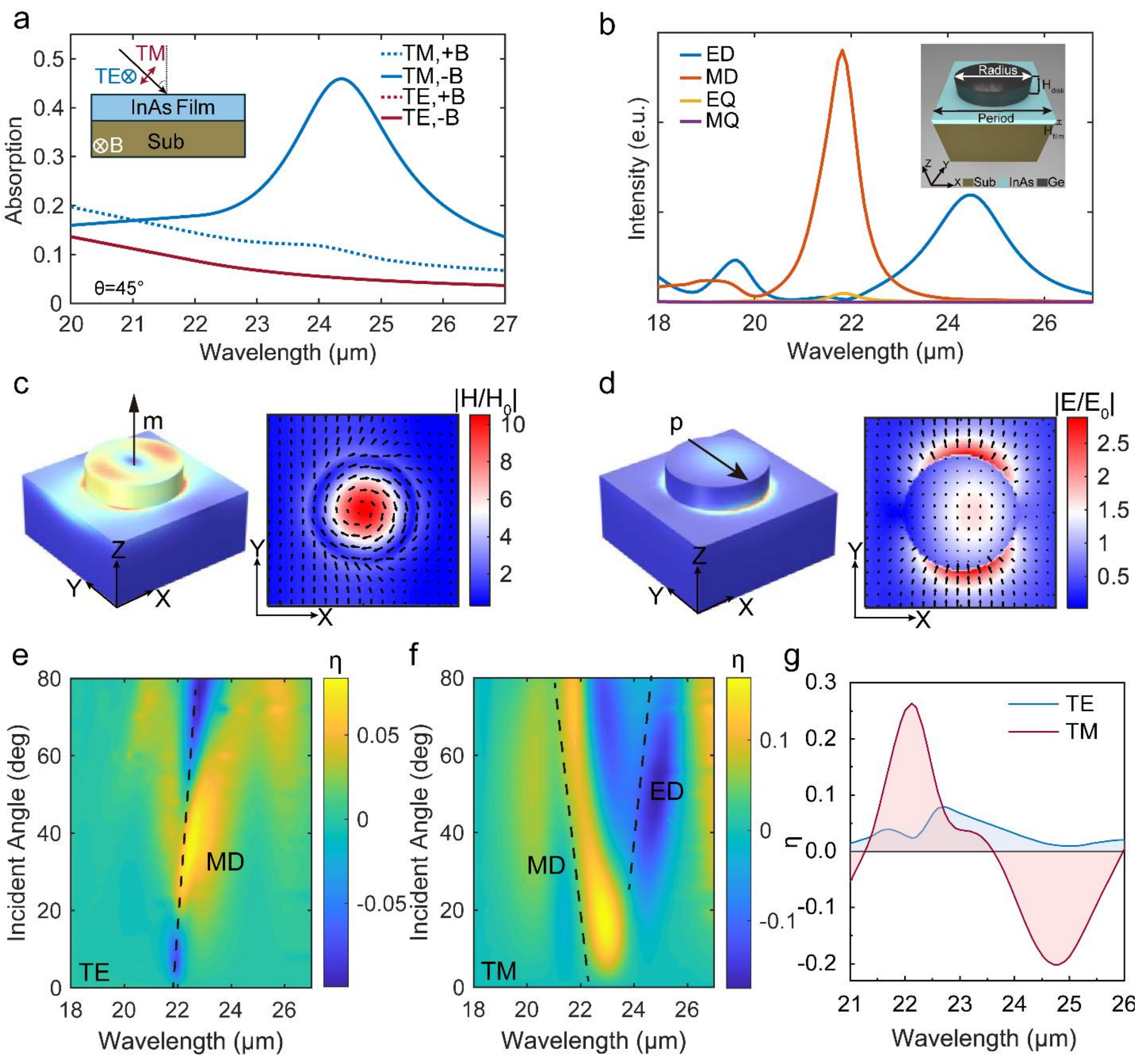

**Fig. 2. Mechanism of breaking polarization constraint in nonreciprocal thermal radiation a,** Absorption spectra of bare ENZ film under two different polarization conditions. **b,** Schematic diagram of the proposed metasurface and the corresponding analytical multipole analysis ($H_{disk}$=2 μm, Radius=3.5 μm, Period=11.5 μm, $H_{film}$=0.7 μm). **c, d,** The near-field distribution in the case of magnetic dipole (MD) and electric dipole (ED) mode under TE polarization. **e, f,** Nonreciprocal absorption spectra changed with incident angles under different polarizations, revealing fundamentally different nonreciprocal response. **g,** Nonreciprocal absorption contrast $\eta$ extracted at an incident angle of θ = 45°, showing a sign reversal between the ED and MD dominated resonances under TM polarization.

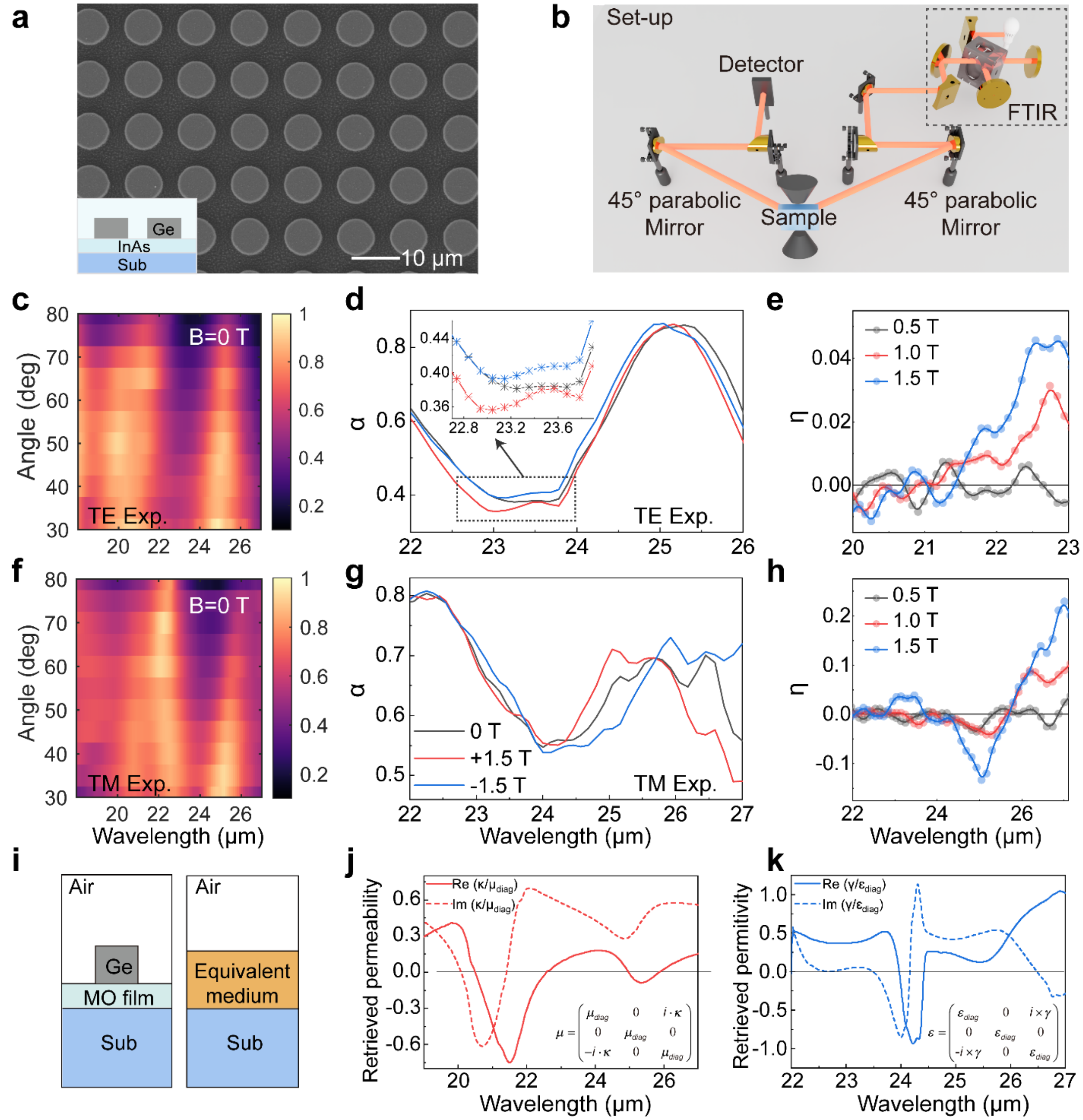

**Fig. 3. Experimental demonstration of dual polarization nonreciprocal absorption. a,** SEM image of the fabricated sample. **b,** Custom-built infrared absorption spectrum characterization setup with a tunable external magnetic field. **c, f,** Measured angle-resolved absorption spectra in the absence of external magnetic field under both TE and TM polarization. **d, g,** Measured nonreciprocal absorption spectra at B=±1.5 T for dual polarization. **e, h,** Change in the absolute absorptivity as a function of magnetic field, extracted near the MD resonance for TE polarization and near the ED resonance for TM polarization, respectively. Each curve is obtained by averaging

three consecutive measurements. **i,** The schematic diagram of resonant structure and its related equivalent material. **j, k,** Off-diagonal components of the complex permeability tensor for TE polarization and permittivity tensor for TM polarization. All results in **d–e** and **g-h** correspond to an incident angle 45°.

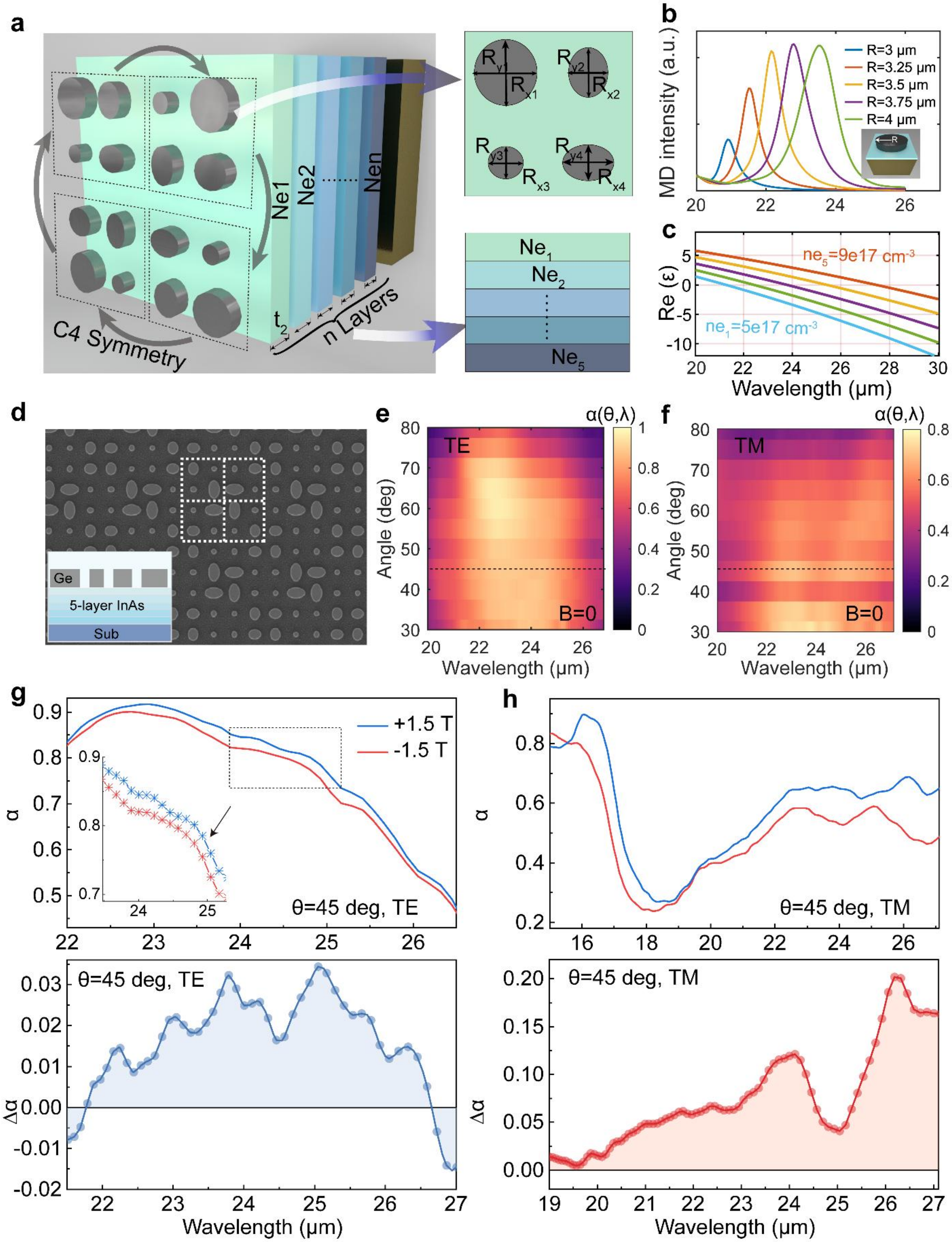

**Fig. 4. Dual polarization control of broadband nonreciprocal absorption. a,** Framework of the optimization process for the broadband design based on genetic algorithm. **b,** MD resonant

contributions of Ge resonators with various dimensions. **c,** Real part of the permittivity of InAs films with gradient doping concentration, revealing that ENZ response can be maintained over a broad spectral range. **d,** SEM image of the fabricated sample corresponding to the optimal design. **e,f,** Measured angle-resolved absorption spectra without magnetic field. **g-h,** Measured nonreciprocal absorption spectra under B= ±1.5 T and the corresponding nonreciprocal contrast over a broad wavelength range at an incident angle θ=45°. **g,** TE polarization. **h,** TM polarization.